\documentclass[5p]{elsarticle}
\biboptions{longnamesfirst,angle,semicolon,sort&compress}
\usepackage{xcolor,colortbl}
\usepackage{graphicx}
\usepackage{amssymb}
\usepackage[amssymb,thinqspace,thinspace]{SIunits}
\usepackage{booktabs}
\usepackage{color}
\usepackage{ctable}
\usepackage{SIunits}
\usepackage{upgreek}
\usepackage[version=3]{mhchem}
\usepackage[modulo,switch]{lineno} 
\usepackage{soul} 
\newcommand{\si}{\SIunits}
\graphicspath{{./figures/}}

\journal{arXiv.org}

\begin{document}
\begin{frontmatter}
\title{Precipitate dissolution during deformation induced twin thickening in a CoNi-base superalloy subject to creep}
\author[str]{Vassili Vorontsov}
\ead{vassili.vorontsov@strath.ac.uk}
\author[ic]{Thomas McAuliffe}
\ead{t.mcauliffe17@imperial.ac.uk}
\author[rr]{Mark C. Hardy}
\ead{mark.hardy@rolls-royce.com}
\author[ic]{David Dye}
\ead{david.dye@imperial.ac.uk}
\author[ic]{Ioannis Bantounas\corref{cor1}}
\ead{ibantounas.ac@gmail.com}
\address[str]{Department of Design, Manufacturing and Engineering Management, University of Strathclyde, James Weir Building, 75 Montrose Street, Glasgow, G1 1XJ, UK}
\address[ic]{Department of Materials, Imperial College, South Kensington, London SW7 2AZ, UK}
\address[rr]{Rolls-Royce plc, Derby, DE24 8BJ, UK}
\cortext[cor1]{Corresponding author at: Hiscox, 1 Great St Helen's, London EC3A 6HX, UK}

\begin{abstract}
The tensile creep performance of a polycrystalline Co/Ni-base superalloy with a multimodal $\upgamma'$ distribution has been examined at \si{800\celsius} and \si{300\,\mega\pascal}. The rupture life of the alloy is comparable to that of RR1000 tested under similar conditions. Microstructural examination of the alloy after testing revealed the presence of continuous $\upgamma'$ precipitates and \ce{M23C6} carbides along the grain boundaries. Intragranularly, coarsening of the secondary $\upgamma'$ precipitates occurred at the expense of the fine tertiary $\upgamma'$. Long planar deformation bands, free of $\upgamma'$, were also observed to traverse individual grains ending in steps at the grain boundaries. Examination of the deformation bands confirmed that they were microtwins. Long sections of the microtwins examined were depleted of $\upgamma'$ stabilising elements across their entire width, suggesting that certain alloy compositions are susceptible to precipitate dissolution during twin thickening. A mechanism for the dissolution of the precipitates is suggested based on the Kolbe reordering mechanism. 

\end{abstract}

\begin{keyword}
Cobalt-base superalloys \sep 
Creep \sep
Microtwinning \sep
Atomic ordering \sep
Transmission electron microscopy



\end{keyword}

\end{frontmatter}


\section{Introduction}

The development of $\upgamma$/$\upgamma'$ two phase Co- and CoNi-based superalloys is promising for improved performance of high temperature structural applications. The exceptional high temperature performance of Ni-based superalloys is attributed to the precipitation of a coherent L1$_2$ ordered $\upgamma'$ phase within a disordered face centred cubic $\upgamma$ matrix. The shearing resistance offered by these ordered precipitates results in the excellent creep resistance they display. 

The discovery of a $\upgamma$/$\upgamma'$ two phase field within the Co-Al-W ternary system \cite{Sato:2006fk,Lee:1971aa} has enabled the development of Co- and CoNi-based superalloys for high temperature applications \cite{Pollock:2010aa, Suzuki:2015fk, Neumeier:2015fk, Bauer:2010aa}. Traditionally, Ni-based superalloys are used for structural applications where a combination of good strength, environmental resistance and creep strength are of importance \cite{Reed:2007bk, Sims:1987bk}. 
Understanding of creep behaviour is critical to the implementation of Co- and CoNi-base superalloys for high performance applications.
Promising single crystal CoNi-base superalloys have been developed with comparable creep resistance to that of 1st generation Ni-base superalloys~\cite{Titus:2012ab,Titus:2012uv}. 

Microtwinning has been observed in Ni-base~\cite{Ardakani:1999aa, Kakehi:2000aa, Knowles:2003aa, Viswanathan:2005aa, Karthikeyan:2006aa, Smith:2016aa} and CoNi-base~\cite{Freund:2017fk} superalloys tested between temperatures of \si{650-750\celsius} and stresses of \si{350-850\,\mega\pascal}. Though twinning is typically associated with high strain rate deformation at low temperatures~\cite{Dieter:1986bk,Christian:1995aa}, these microtwins form at intermediate temperatures under slow strain rates. Knowles and Chen~\cite{Knowles:2003aa} have suggested that microtwins nucleate from superlattice extrinsic stacking faults (SESFs).

Several mechanisms have been suggested for the creation of such faults~\cite{Kear:1969aa,Condat:1987aa,Decamps:1991ab,Decamps:1993aa,Kolbe:2001aa,Knowles:2003aa,Decamps:2004aa}, the reader is referred to~\cite{Unocic:2008uv} for a comprehensive review. 
However, diffraction contrast experiments~\cite{Viswanathan:2005aa} and high resolution TEM micrographs~\cite{Koravik:2009aa,Freund:2017fk} have confirmed the presence of $\frac{1}{6}$$<$$112$$>$ dislocations at the leading edge of extended stacking faults on adjacent $\{111\}$ planes. Such an observation would require dislocation decorrelation~\cite{Unocic:2011uv} and suggests that the likely twinning mechanism is that suggested by Kolbe~\cite{Kolbe:2001aa}. Shearing of the L1$_2$ superlattice by $\frac{1}{6}$$<$$112$$>$ Shockley partials results in the formation of a pseudo twin~\cite{Christian:1988aa}. In such a configuration the order of the L1$_2$ superlattice is not preserved across the twin boundary. Kolbe~\cite{Kolbe:2001aa} suggested that a ``true'' twin structure could be restored via a two step atomic re-ordering process involving the short range diffusion of atomic species in the wake of the leading partials. 

In addition to short range atomic re-ordering, long range diffusion of $\upgamma$ stabilising elements have been shown to segregate to the leading partials~\cite{Smith:2017aa,Barba:2017aa,Barba:2017ab}. This Cottrell atmosphere is thought to lower the energy penalty associated with forming a complex stacking fault (CSF) when shearing the $\upgamma'$ via $\frac{1}{6}$$<$$112$$>$ partials~\cite{Rao:2018aa} and is considered as the rate limiting step during twin formation~\cite{Smith:2017aa}. Segregation of $\upgamma$ stabilising elements is not restricted to the leading partials but has also been observed at the fault boundaries~\cite{Viswanathan:2015aa,Titus:2015aa,Eggeler:2016fk,Freund:2017fk}, an observation not limited to the Ni- and CoNi-base alloy systems but having also been reported in MnAl Heusler alloys~\cite{Palanisamy:2018aa} and Mg alloys~\cite{Nie:2013aa}.
In some alloy systems, segregation of elements to stacking faults can stabilise new phases within the fault~\cite{Smith:2015aa,Titus:2015aa,Titus:2016aa,Makineni:2018ac}. Examples include the $\upeta$ phase in a variant of the Ni-base superalloy ME3~\cite{Smith:2015aa} and the DO$_{19}$ phase in the Co-Al-W-Ta system~\cite{Titus:2016aa}.

Precipitation of carbides after prolonged high temperature exposure has been reported in polycrystalline and single crystal Ni-base superalloys~\cite{Garosshen:1985aa,Rao:1983aa,Sundararaman:1997aa,Wu:2008aa,Xu:1998aa}. In Cr rich alloys aged at intermediate temperatures, \ce{M23C6} carbides precipitate with a cube-cube orientation relationship with the parent matrix~\cite{Chen:2002aa,Wu:2008aa}. Carbides are commonly associated with regions of localised deformation and, together with grain boundaries, are thought to be dislocation nucleation sites \cite{Unocic:2011uv,Phillips:2010fk,Phillips:2013fk}. In the present work, evidence suggests such carbides to be the sources of twinning.

In this paper a $\upgamma/\upgamma'$ CoNi-base superalloy with a bimodal $\upgamma/\upgamma'$ size distribution has been creep tested to rupture at \si{800\celsius} and \si{300\,\mega\pascal}. Examination of the post-test microstructure revealed the presence of precipitate-free bands which traversed entire $\upgamma$ grains. These bands were confirmed to be deformation-induced twins which are thought to nucleate at the incoherent interface of grain boundary \ce{M23C6} carbides. It is postulated that the ordered precipitates dissolve into the $\upgamma$ matrix during thickening of the twin. A mechanism for the dissolution of the precipitates is proposed and the consequences of these structures discussed within the text. To our knowledge, this is the first report of precipitate dissolution during twin growth in superalloys.

\section{Experimental}

A CoNi-base superalloy designated V208C, developed by Knop~\emph{et al.}~\cite{Knop:2014fk}, was melted in a vacuum arc-melter backfilled with argon and poured into a \si{23$\times$23$\times$60\,\milli\meter} mould. 
Table~\ref{tab:alloy-comp} shows the nominal alloy composition.
The ingot was homogenised in a vacuum furnace at \si{1250\celsius} for \si{48\,\hour} and profile rolled at \si{1200\celsius} down to a \si{13$\times$13\,\milli\meter} cross cross section with a maximum of 15\% strain per pass. Samples were machined into creep specimens with a \si{14.8\,\milli\meter} gauge length and \si{5\,\milli\meter} gauge diameter. 
Annular ridges were machined on enlarged shoulders of the test specimen for attaching an extensometer.
Creep specimens were encapsulated in quartz tubes under low pressure argon and solution treated at \si{1100\celsius} for \si{1\,\hour}, slow cooled at \si{20\celsius\per\minute} to \si{700\celsius} and air cooled (AC) to room temperature (RT). The specimen was further aged at \si{825\celsius} for \si{4\,\hour} followed by AC to RT.

\newcolumntype{g}{>{\columncolor{lightgray}}c}
\ctable[
    caption  = {Nominal compositions (at. \%) and $\upgamma'$ area fraction of alloy V208C (Patent Ref.~\cite{Dye:2017pat}).}, 
    label    = tab:alloy-comp,
    doinside = \scriptsize,
    pos      = hbtp ]{@{}ccccccccc|c@{}}{
 }{                                                          \FL
        Co & Ni & Cr & Al & W & Ta & C & B & Zr & $\upgamma$' fract. \ML 
        36.1&34&15&10.5&3&1& 0.15&0.2&0.04& 52\% \LL 
}

The grain size of the alloy was \si{33$\pm$4\,\micro\meter}, including twins, as measured from optical micrographs using the Abrams three circle method outlined in ASTM E112.32609. Average equivalent area diameter of the secondary and tertiary $\upgamma'$ precipitates were measured using ImageJ software. The $\upgamma_s'$ and $\upgamma_t'$ sizes were \si{100 and 22\,\nano\meter}, respectively. 

Creep tests were performed in tension on a lever-loading machine under a constant force to rupture\footnote{The test was paused and restarted after \si{360\,\hour} (\si{$\approx$16\%} strain) as the displacement limit of the test apparatus was reached. This involved removing the load from the specimen, cooling to room temperature and re-adjusting the aparatus prior to restarting the test. At this point, the material was within the tertiary creep regime with \si{30\,\hour} remaining to rupture.}. The test temperature was set to \si{800\celsius} and the starting stress to \si{300\,\mega\pascal}. Two thermocouples attached with high temperature wire to the centre of the gauge were used to monitor temperature during testing. Creep strain was recorded by fitting an extensometer to the annular ridges on the test piece. The change in displacement was measured using a transducer attached to a pair of arms extending below the furnace. 

The deformed microstructure was examined by sectioning the gauge perpendicular to the loading direction. Samples were progressively ground down to 4000 grit SiC paper and then polished using a suspension containing 100ml water, 30ml coloidal silica suspension and 70ml H$_2$O$_2$. The $\upgamma'$ precipitates were revealed by electroetching in a solution of 2.5\% phosphoric acid in methanol using a voltage of 10V for 10 seconds. Transmission electron microscopy (TEM) lamellae were prepared from the polished cross-section using an FEI Helios Nanolab 600 focussed ion beam workstation via the \emph{in-situ} lift-out technique~\cite{Giannuzzi:1999ab,Mayer:2007uq}. The twin chemistry was examined in a JEOL JEM-2100F microscope equipped with a XEDS silicon-drift detector from Oxford Instruments. XEDS data was collected and analysed using the Oxford Instruments AZTec v3.1 software package. Electron micrographs were acquired with the InLens detector on Zeiss Sigma 300 and a Zeiss Auriga Cross-Beam scanning electron microscopes. The secondary electron images presented were acquired on a FEI Helios Nanolab 600 workstation.

\section{Results}
\subsection{Creep Behaviour}

\begin{figure}[htbp]
	\centering
	\includegraphics[width=0.48\textwidth]{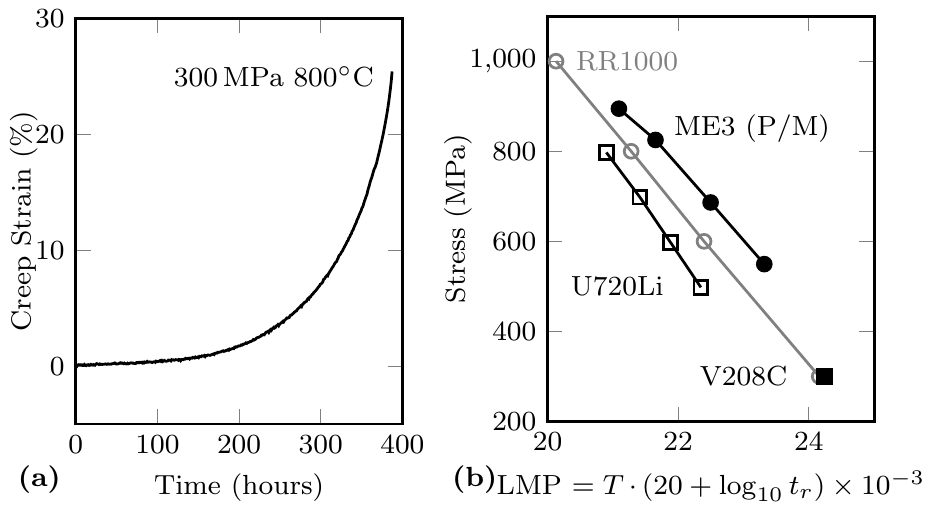} 
	\caption{(a) Creep curve of specimen tested in tension at \si{800\celsius} and \si{300\,\mega\pascal}, and (b) comparison of creep performance with U720Li~\cite{Gu:2009aa}, RR1000~\cite{Christofidou:2018aa} and ME3~\cite{Gabb:2002aa} on a Larson Miller parameter graph.}
	\label{fig:mech-prop}
\end{figure}

The tensile creep strain vs time curve for the alloy tested at \si{800\celsius} and \si{300\,\mega\pascal} is shown in Figure~\ref{fig:mech-prop}~(a). The time to rupture was \si{390\,\hour} with a minimum creep rate of \si{8.6E-9\,\reciprocal\second}. Figure~\ref{fig:mech-prop} (b) shows the test results displayed on a Larson-Miller parameter (LMP) plot, equation~\ref{eq:lmp}. 

\begin{equation}
	\text{LMP} = T \cdot (20 + \log_{10} t_r) \times 10^{-3} \, ,
	\label{eq:lmp}
\end{equation}

where $T$ is the test temperature in Kelvin and $t_r$ the time to rupture in hours. 

The graph compares the current alloy to the performance of industrially relevant disk alloys U720Li~\cite{Gu:2009aa}, RR1000~\cite{Christofidou:2018aa} and ME3 (P/M)~\cite{Gabb:2002aa}. At a stress of \si{300\,\mega\pascal} V208C in the current microstructural condition has a comparable LMP to that of RR1000~\cite{Christofidou:2018aa}, a disk superalloy currently used in gas turbine engines.

\subsection{Grain Boundary Evolution}
\label{gb-evo}

\begin{figure*}[htbp]
	\centering
	\includegraphics[width=1.00\textwidth]{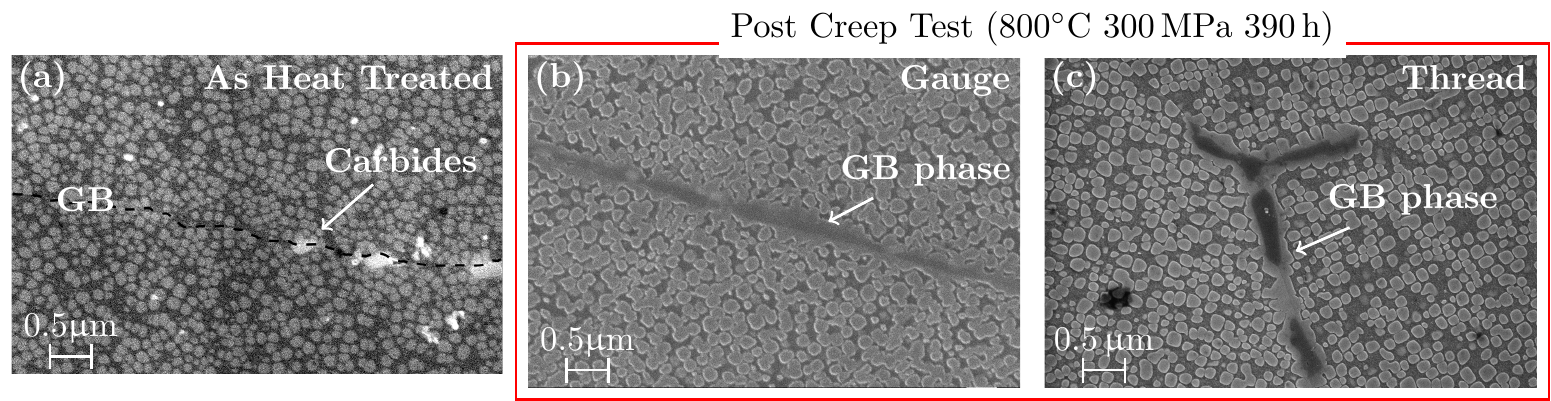}  
	\caption{InLens micrographs showing (a) grain boundary carbides in the pre-test microstructure, (b) and (c) growth of continuous grain boundary phases post creep test at \si{800\celsius} for \si{390\,\hour} in the specimen gauge and thread, respectively.}
	\label{fig:micro-gb}
\end{figure*}

Exposing the alloy to a temperature of \si{800\celsius} for \si{390\,\hour} precipitated out Cr rich \ce{M23C6} carbides and $\upgamma'$ at the $\upgamma$ grain boundaries. The grain boundary structure before creep testing is shown in Figure~\ref{fig:micro-gb}~(a). A dashed line is drawn on the micrograph to outline the boundary, as there is little contrast to distinguish it from the bulk of the two adjacent grains. Few carbides can be seen decorating the boundary. The post creep test microstructure is shown in Figures~\ref{fig:micro-gb}~(b) and (c). Micrographs from the gauge section after exposure at \si{800\celsius} and \si{300\,\mega\pascal} for \si{390\,\hour} reveal the presence of phases decorating the grain boundaries of the alloy, Figure~\ref{fig:micro-gb}~(b). These phases are also observed within the microstructure from the threads of the specimen, Figure~\ref{fig:micro-gb}~(c), implying that they are not stress induced phases but rather a result of heat treatment during testing.

The grain boundary phases which formed after testing are Cr rich \ce{M23C6} carbides and the L1$_2$ $\upgamma'$ phase. Figure~\ref{fig:ctem-gb}~(a) shows a centre beam dark field (CBDF) conventional (C)TEM micrograph of a foil prepared from the gauge of the creep specimen. Two grains labelled G1 and G2 are visible in the top left and bottom right of the micrograph, respectively. Grain G2 is oriented along the $[001]$ zone axis, Figure~\ref{fig:ctem-gb}~(c), with a superlattice reflection selected for the CBDF micrograph. The $\upgamma'$ precipitates appear bright against a dark $\upgamma$ matrix, while grain G1 does not contribute to the image and appears dark. A carbide is located along the grain boundary, more clearly visible in the centre beam dark field (CBDF) micrograph in Figure~\ref{fig:ctem-gb}~(b). The carbide is oriented along the $[001]$ zone axis and its crystal structure resembles that of \ce{M23C6}, Figure~\ref{fig:ctem-gb}~(d).

Scanning transmission electron microscopy (STEM) energy dispersive spectroscopy (EDS) maps reveal that the carbide from Figure~\ref{fig:ctem-gb}~(b) is enriched in Cr, Figure~\ref{fig:ctem-gb}~(f). Additionally, the carbide is surrounded by $\upgamma'$ precipitates and bound between two Ta/Zr rich carbides. The compositions of the Cr and Ta/Zr-rich carbides are presented in Table~\ref{tab:carbide-comp}.

\newcolumntype{g}{>{\columncolor{lightgray}}c}
\ctable[
    caption  = {Carbide compositions (at. \%) measured via STEM-EDS.}, 
    label    = tab:carbide-comp,
    doinside = \scriptsize,
    pos      = hbtp ]{@{}lrrrrrrrr@{}}{
 }{                                                          \FL
	Phase       &  Co & Ni &  Al & W &  Ta &  Cr &  C & Zr \ML
	\ce{M23C6}         &   6 &   2 &   0 &  3 &   0 &  75 &  13 &  0 \NN
	\ce{MC}         &   5 &   2 &   0 &  4 &  24 &  31 &  26 &  9 \LL
}

The Cr enrichment coupled with the large cubic lattice parameter suggests that this is an \ce{M23C6} carbide with approximate composition \ce{Cr23C6}. The space group of this phase is classified as $Fm\bar3m$ in Hermann-Mauguin notation. In turn, Ta/Zr-rich carbides tend to be MC carbides. The formation of \ce{M23C6} and $\upgamma'$ at the grain boundaries after prolonged exposure to high temperature is due to the decomposition of MC grain boundary carbides according to the $\text{MC}+\upgamma\rightarrow\text{\ce{M23C6}}+\upgamma'$ reaction sequence~\cite{Reed:2007bk,Sims:1987bk}. 

\subsection{Carbide Induced Slip}
\begin{figure*}[htbp]
	\centering
	\includegraphics[width=0.84\textwidth]{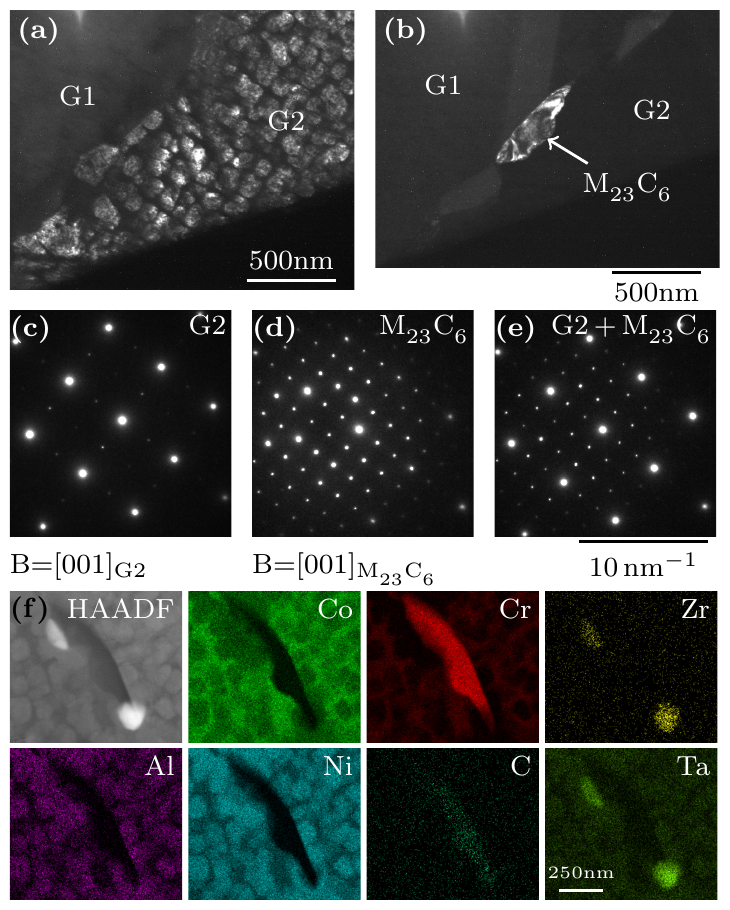} 
	\caption{CBDF micrographs from grain G2 showing (a) intragrannular and grain boundary $\upgamma'$ precipitates and (b) grain boundary \ce{M23C6} carbide coherent with grain G2. SADPs from (c) the bulk of grain G2 and (d) the \ce{M23C6} carbide. (e) SADP showing the cube-cube orientation relationship between G2 and the \ce{M23C6} carbide. All SADPs imaged along the $[001]$ beam direction. (f) STEM HAADF micrograph of carbide with accompanying EDS maps. All micrographs acquired from gauge of creep specimen tested at \si{300\,\mega\pascal} and \si{800\celsius} for \si{390\,\hour}.}
	\label{fig:ctem-gb}
\end{figure*}

\begin{figure}[htbp]
	\centering
	\includegraphics[width=0.40\textwidth]{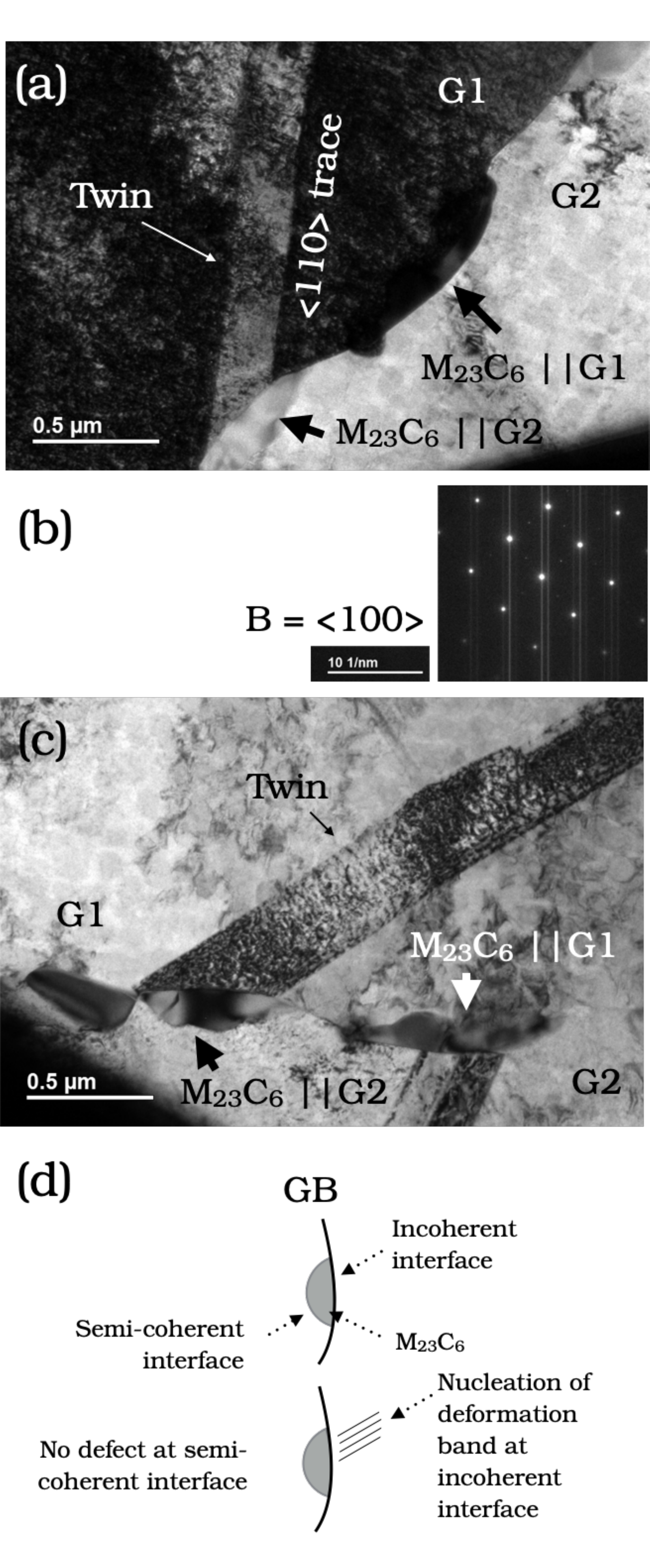} 
	\caption{(a) BF TEM micrograph taken along the $<$$100$$>$ zone axis of grain G1 and carbide \ce{M23C6}\textbar\textbar G1 showing deformation twin intersect the foil plane along the $<$$110$$>$ trace. (b) SADP from (a) - Note that the vertical streaks in the DP are due to charging of the CCD camera and do not reflect a property of the material. (c) BF TEM micrograph showing deformation twin in G1 ending at grain boundary carbide labelled \ce{M23C6}\textbar\textbar G2 which shares a cube-cube relationship with grain G2 and an incoherent interface with G1. (d) Schematic showing defect nucleation at an incoherent interface as a result of grain boundary \ce{M23C6} carbide precipitation and growth.}
	\label{fig:gb-slip}
\end{figure}

Our results show that the growth of \ce{M23C6} follows a cube-cube relationship with one of the grains while inducing strain and the nucleation of slip bands in the adjacent grain, Figure~\ref{fig:gb-slip}. A selected area diffraction pattern (SADP) of grain G2 imaged along the $[001]_{\upgamma}$ zone is shown in Figure~\ref{fig:ctem-gb}~(c), a SADP of the \ce{M23C6} carbide which shares a cube-cube relationship with this grain is shown in Figure~\ref{fig:ctem-gb}~(d). A SADP of both the carbide and $\upgamma$ grain are presented in Figure~\ref{fig:ctem-gb}~(e). These diffraction patterns clearly display the orientation relationship between the carbide and matrix. A CBDF micrograph of the carbide which shares a relationship with grain G2 is provided in Figure~\ref{fig:ctem-gb}~(b). This carbide is labelled as ``\ce{M23C6}\textbar\textbar G2'' in Figure~\ref{fig:gb-slip}~(a) and (c), denoting it's cube-cube orientation relationship with grain G2, and is connected to a deformation twin in the adjacent grain G1. Likewise, carbide ``\ce{M23C6}\textbar\textbar G1'' shares a cube-cube relationship with grain G1 and is connected to a deformation band in the adjacent grain G2. 

We can infer that the carbides nucleate and form a semi-coherent interface with the grain they share an orientation relationship with while exerting a strain on the adjacent grain with which they share an incoherent interface during growth. Coupled with the load from the creep test, these incoherent interfaces make for favourable sites for the initiation of slip events. Delaying the precipitation of these carbides could thus prolong the creep life of the alloy. 

\subsection{Microstructural Evolution}

\begin{figure}[htbp]
	\centering
	\includegraphics[width=0.40\textwidth]{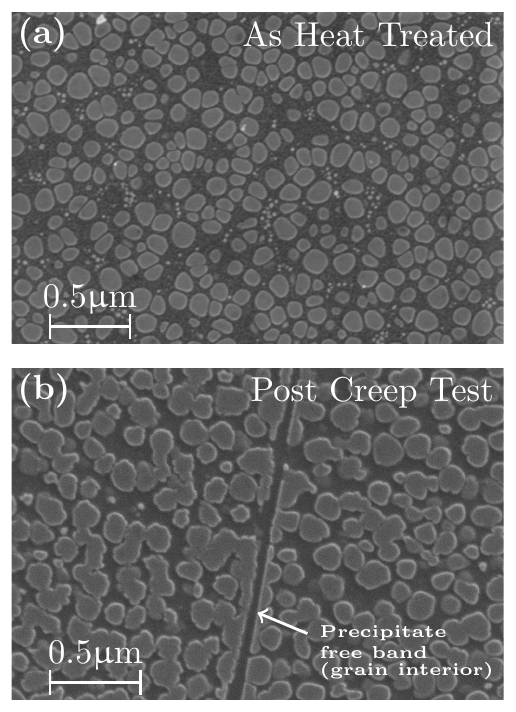} 
	\caption{InLens micrographs of (a) the as-heat treated microstructure showing the secondary and tertiary $\upgamma'$ precipitates, and (b) post-test microstructure from the gauge showing coarsening and coalescence of secondary $\upgamma'$ along with precipitate-free bands within the grain interior.}
	\label{fig:micro-ppt}
\end{figure}

A bimodal $\upgamma'$ microstructure consisting of coarse secondary and finer tertiary precipitates was obtained after heat treating the rolled ingot. The microstructure is presented in Figure~\ref{fig:micro-ppt}~(a). It consists of secondary $\upgamma'$ precipitates with a mean diameter of \si{100\,\nano\meter} and \si{22\,\nano\meter} tertiary precipitates. The seconary and tertiary precipitate area fractions were 49 and 3\%, respectively. After creep testing the secondary $\upgamma'$ precipitates coarsen and coalesce at the expense of the tertiary, Figure~\ref{fig:micro-ppt}~(b). In addition, bands free of precipitates which traverse entire $\upgamma$ grains are observed. These bands, later shown to be deformation microtwins, are the focus of subsequent sections.

\subsection{Deformation Microstructure}

\begin{figure}[htbp]
	\centering
	\includegraphics[width=0.40\textwidth]{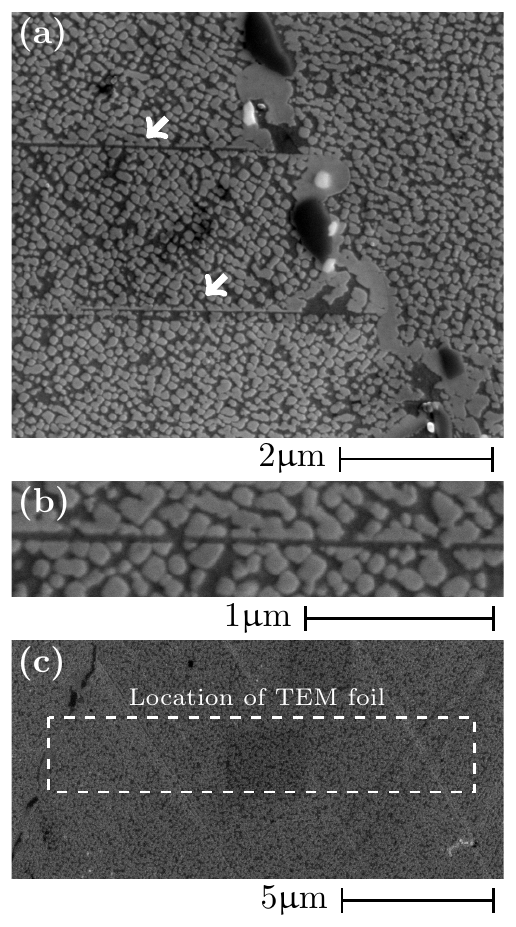} 
	\caption{Secondary electron micrographs showing (a) precipitate free bands (white arrows) terminating at a grain boundary, (b) magnified view of precipitate free bands from the grain interior, and (c) location of TEM foil encompassing multiple bands.}
	\label{fig:micro-bands}
\end{figure}

Figure~\ref{fig:micro-bands}~(a) shows a section of the microstructure of the crept specimen taken from the gauge perpendicular to the loading direction. The arrows indicate precipitate free bands. These bands end at grain boundary steps. Such steps are responsible for the nucleation of grain boundary cavities leading to tertiary creep. Grain boundary $\upgamma'$ and Cr rich \ce{M23C6} carbides, characterised in the previous section, can also be seen to decorate the grain boundary. 

A magnified view of the grain interior showing a deformation band is displayed in Figure~\ref{fig:micro-bands} (b). The figure highlights the abscence of precipitates within the band. In order to characterise the deformation and chemical structure of the band, a TEM foil was extracted from a deformed grain. Figure~\ref{fig:micro-bands}~(c) shows the location of the TEM lamella prior to FIB milling. The area was selected in order to encompass multiple bands within the foil.

\begin{figure*}[htbp]
	\centering
	\includegraphics[width=1.00\textwidth]{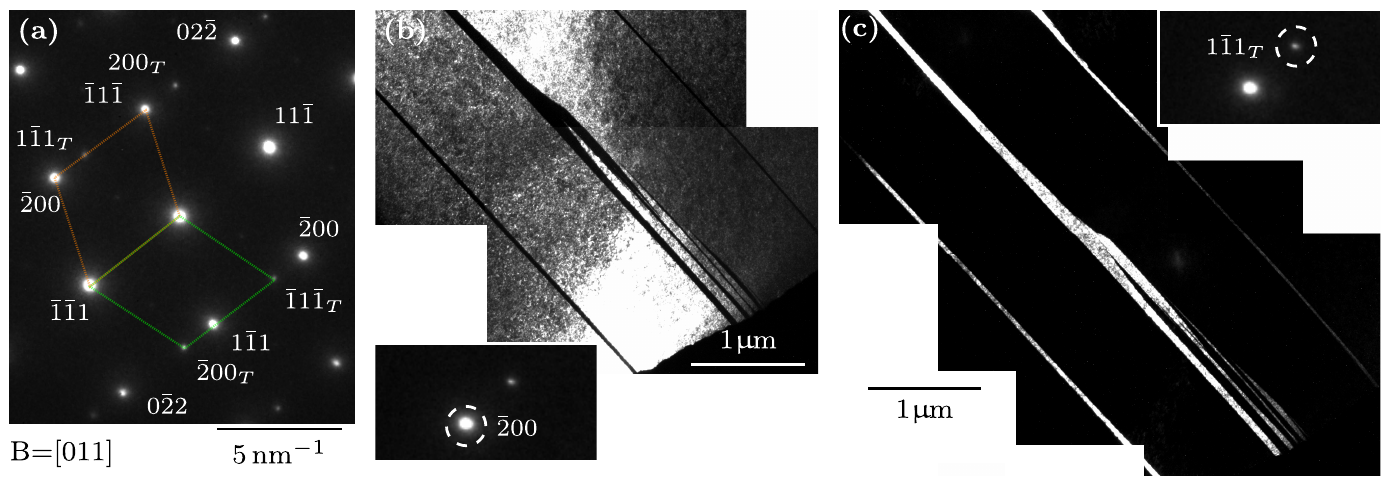} 
	\caption{(a) Selected area diffraction pattern along the $[110]$ zone showing twin reflections (marked with a `T' subscript), and centre beam dark field micrographs from (b) matrix $[\bar200]$ beam with twins appearing dark and (c) twin $[1\bar11]_T$ reflection with twins appearing bright.}
	\label{fig:ctem}
\end{figure*}

Conventional transmission electron microscopy reveals that these deformation bands are deformation twins on the order of a few hundred nanometers in thickness. A selected area diffraction pattern (SADP) from an area containing the precipitate free bands is shown in Figure~\ref{fig:ctem} (a). The foil is oriented such that the beam direction is along the $[011]$ zone axis of the grain. Along this zone, diffuse diffraction spots can be seen about the $(\bar1\bar11)$ plane, revealing the twin nature of the precipitate free bands.

A $[\bar200]$ matrix reflection was selected to create a centre beam dark field (CBDF) micrograph in Figure~\ref{fig:ctem} (b). The parent grain is bright and the deformation bands dark. Figure~\ref{fig:ctem} is a CBDF micrograph using a $[1\bar11]$ twin reflection. In this instance the parent grain appears dark and the twin structure bright. These micrographs confirm that the deformation bands observed in the SEM are deformation twins. Consistent with the absence of $\upgamma'$ precipitates observed in the SEM micrographs, no superlattice reflections can be observed from the twinned reflections. Despite the lack of superlattice reflections the chemical compsition of the twin suggests regions of high concentration of $\upgamma'$ stabilising elements.

\subsection{Twin Chemistry}

\begin{figure*}[htbp]
	\centering
	\includegraphics[width=0.84\textwidth]{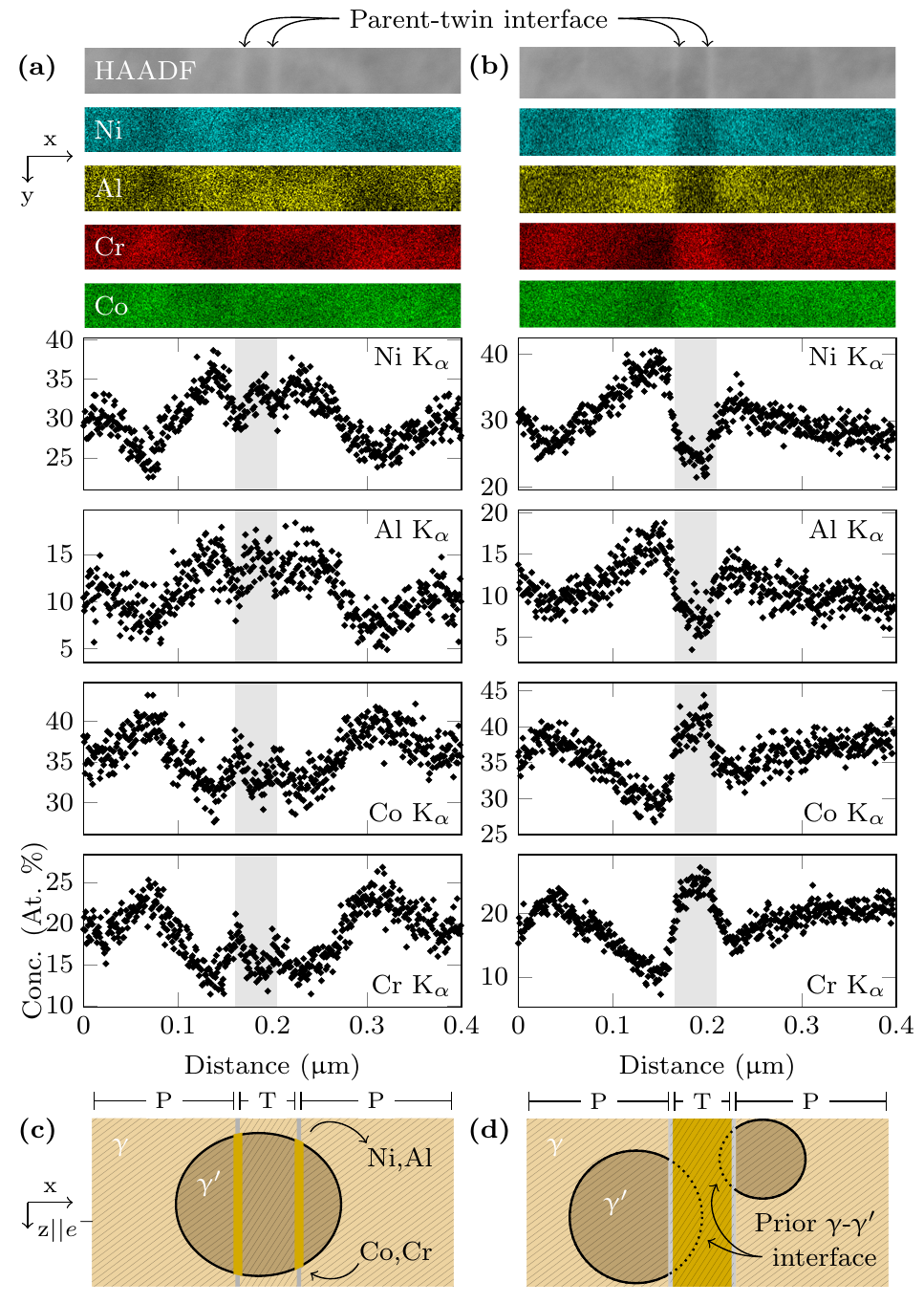} 
	\caption{X-ray line profiles across thickness of twin showing (a) enrichment of $\upgamma$ and depletion of $\upgamma'$ stabilising elments at twin/matrix interface when the twin is confined within a $\upgamma'$ precipitate, (b) consumption of $\upgamma'$ precipitates by $\upgamma$ stabilising elements during twin thickening. Schematics (c) and (d) show the physical representations of the line profile observations from (a) and (c), respectively. Hashed lines represent the ($\bar101$) planes. `P' and `T' in the schematics denote regions of the parent and twin, respectively.}
	\label{fig:stem}
\end{figure*}

Consistent with the majority of observed deformation twins from SEM micrographs the twin channels are enriched in $\upgamma$ stabilising elements, as shown via STEM-XEDS maps in Figure~\ref{fig:stem}(b). The X-ray maps show the distribution of Co, Cr, Ni and Al within two twinned regions. In this alloy, Co and Cr partition to the $\upgamma$ phase while Ni and Al partition to the $\upgamma'$ phase~\cite{Knop:2014fk}. Despite the SEM micrographs revealing an absence of $\upgamma'$ precipitates within the deformation channels, the elemental maps show concentrations of $\upgamma'$ stabilising elements coupled with regions rich in $\upgamma$ stabilising elements. The composition of elements across the twin in both such cases is examined next.

Line profiles across the thickness of the twin are presented from regions rich in $\upgamma'$ and $\upgamma$ stabilising elements, Figure~\ref{fig:stem}(a) and (b), respectively. We first examine the case of a twinned region containing an $\upgamma'$ precipitate, or more specifically, a region rich in Ni and Al $\upgamma'$ stabilising elements, Figure~\ref{fig:stem}(a). Tracking the Ni K$_{\alpha}$ quantitative line profile from left to right shows a decrease in concentration to a minimum of \si{$\approx$ 24\,at.\%} followed by a rise to \si{$\approx$ 36\,at.\%}. This first dip is believed to be due to the converged electron beam crossing from a $\upgamma + \upgamma'$ region of the foil to a $\upgamma$ channel, the composition of which is consistent with atom probe tomography results of the $\upgamma$ phase in this alloy~\cite{Knop:2014fk}. The beam then scans across a $\upgamma + \upgamma'$ region of the foil, which becomes progressively richer in $\upgamma'$. Following the peak in Ni composition of \si{$\approx$ 36\,at.\%} there is a dip in concentration of \si{$\approx$ 6\,at.\%} Ni at the twin/matrix interface. The Ni concentration increases throughout the thickness of the twin to \si{$\approx$ 34\,at.\%}, however does not recover to the composition of the undeformed precipitate. A second dip in the concentration is again observed at the adjacent twin/matrix interface. After the electron beam crosses this planar interface the Ni concentration gradually decreases from a $\upgamma + \upgamma'$ phase region to that of pure $\upgamma$. The Al K$_{\alpha}$ line profile follows a similar trend to that of Ni, both elements preferentially partitioning to the $\upgamma'$ phase. 

Line profiles of X-ray peaks from the $\upgamma$ stabilisers, Co K$_\alpha$ and Cr K$_\alpha$, follow an inverse pattern to that of Ni and Al. The concentration of Co and Cr gradually decreases as the electron beam moves from the $\upgamma$ channel to a region containing $\upgamma + \upgamma'$ phases. At the twin/matrix interfaces the concentration of $\upgamma$ stabilising elements increases. A schematic of the precipitate and twin configuration is depicted in Figure~\ref{fig:stem}(c) where a twin is confined within a $\upgamma'$ precipitate with its twin/matrix interfaces depleted in $\upgamma'$ and enriched in $\upgamma$ stabilising elements. This is attributed to long range diffusion of $\upgamma$ stabilising elements during twin lengthening through a $\upgamma'$ precipitate~\cite{Barba:2017aa,Barba:2017ab} and helps to lower the energy penalty associated with the passage of $\frac{1}{6}$$<$$112$$>$ partials on adjacent $\{111\}$ planes during twin thickening~\cite{Freund:2017fk}.

A different scenario is presented in Figure~\ref{fig:stem}(b) where the composition across the entire thickness of the twin resembles that of the $\upgamma$ phase. Tracking the Ni K$_\alpha$ line profile. The concentration of Ni gradually increases from \si{$\approx$ 26\,at.\%}, in the $\upgamma$ phase, to \si{$\approx$ 39\,at.\%} as it enters a $\upgamma + \upgamma'$ phase region. A sudden drop to \si{$\approx$ 26\,at.\%} is observed at the twin/matrix interface. Up to this point, this observation is consistent with the previous case when the twin is contained within a $\upgamma'$ precipitate. However, unlike the previous example, the concentration of Ni does not recover to that of the $\upgamma'$ precipitate but remains low at \si{$\approx$ 25\,at.\%}, resembling that of the $\upgamma$ phase, throughout the thickness of the twin. As the converged beam crosses the adjacent twin/matrix interface the concentration of Ni sharply rises again by \si{$\approx$ 5\,at.\%} and then follows a gradual increase to \si{$\approx$ 34\,at.\%} before gradually decreasing to \si{$\approx$ 27\,at.\%}. The raised Ni concentration on either side of the twin is thought to be two separate $\upgamma'$ precipitates. The Al K$_\alpha$ peak follows a similar trend to that of Ni, while the Co and Cr K$_\alpha$ peaks, being $\upgamma$ stabilisers, follow an inverse pattern. I.e., deplete when the electron beam scans the $\upgamma'$ precipitate and increase in the twin region.

In this instance we believe that the twin has nucleated (i.e. lengthened) within the $\upgamma$ matrix and consumed the adjacent $\upgamma'$ precipitates during the thickening stage of its life. A schematic is shown in Figure~\ref{fig:stem}(d) to depict the migration of $\upgamma$ stabilising elements to the prior precipitate boundary.  This creates a region along the length of the twin depleted in $\upgamma'$ stabilising elements with a composition very close to that of the $\upgamma$ phase which stretches across the entirity of its width. A mechanism is proposed in the disussion for the creation of such a twinned region.

\section{Discussion}

\subsection{Carbide precipitation and slip}

The precipitation of \ce{M23C6} carbides at the grain boundaries introduces strain in the adjacent grain which, coupled with the load from the creep test, make the carbides favourable areas for slip nucleation. The work from Section~\ref{gb-evo} clearly shows grain boundary \ce{M23C6} carbides precipitating coherently with one of the grains adjacent to the grain boundary. Figure~\ref{fig:gb-slip}(c) shows two \ce{M23C6} carbides precipitating coherently with one of the grains and inducing slip on the adjacent grain. Carbide \ce{M23C6}\textbar\textbar G1 shares a cube-cube orientation relationship with grain G1. The carbide shares a semi-coherent interface with this grain while its interface with the adjacent grain G2 is incoherent. A deformation twin is seen to lie on the adjacent grain G2 intersecting grain G1 at the incoherent interface of the \ce{M23C6}\textbar\textbar G1 carbide. The same can be observed for carbide \ce{M23C6}\textbar\textbar G2 which has a cube-cube orientation relationship with G2 and induces slip in the adjacent grain G1. A schematic depicting these observations is presented in Figure~\ref{fig:gb-slip}(d).

In support of the above argument, SEM micrographs from the threads of the creep specimen reveal that stress is not a prerequisite for the formation of these grain boundary carbides, Figure~\ref{fig:micro-gb}~(c). This means that in trying to understand the spacial coincidence of the carbide and deformation structure we are lead to believe that the defects are not a prerequisite for carbide nucleation but that likely the opposite is true. Carbides precipitate at the grain boundaries due to thermodynamics and act as nucleation sites for the deformation structures.

The implication of this observation is that there could be room to improve the creep performance of this alloy by lowering the Cr additions. Grain boundary \ce{M23C6} is stabilised by chromium. Reducing chromium could suppress the formation of the grain boundary carbide during exposure to high temperature. If the carbides indeed act as slip initiation sites for twin formation, as discussed above, their suppression could also delay the onset of such deformation. This in turn would prolong the time to rupture and improve the creep performance of the alloy. Given the comparable performance of this alloy with RR1000~\cite{Christofidou:2018aa}, a Ni-base superalloy in current use in jet engines, there is therefore potential for polycrystalline CoNi-base superalloys to outperform current generation disk Ni-base superalloys.

\subsection{Twinning mechanism}

In this study we are concerned with twin growth and how it relates to creep rupture. We observe long regions of precipitate-free twins which give rise to grain boundary steps, believed to be the cause for the transition to the tertiary creep regime and subsequently rupture. Two mechanisms are proposed for twin growth based on the presence of $\upgamma'$ precipitates within the vicinity of the original twin. Both cases can be accounted for by the following example. 

A single twin is nucleated from a SF that has grown across both the $\upgamma$ and $\upgamma'$ phases from one end of a grain to the other. Across the length of the fault both phases have undergone a displacement. As the twin grows perpendicular to the $\{$$111$$\}$ twin plane it carries with it the segregation of $\upgamma$ stabilising elements. When the twin length is confined to one $\upgamma'$ precipitate, Figure~\ref{fig:stem}(c), the twin boundaries are enriched in Co and Cr and depleted in Ni and Al. In instances where the twin has nucleated within a precipitate and continues to grow into the $\upgamma$ matrix, once the twin boundary reaches the adjacent precipitate instead of cutting into the precipitate, dissolution of the precipitate occurs, Figure~\ref{fig:stem}(d). The twin boundary consumes the neighbouring precipitate. 

\subsection{Proposed precipitate dissolution mechanism}

\begin{figure}[htbp]
	\centering
	\includegraphics[width=0.48\textwidth]{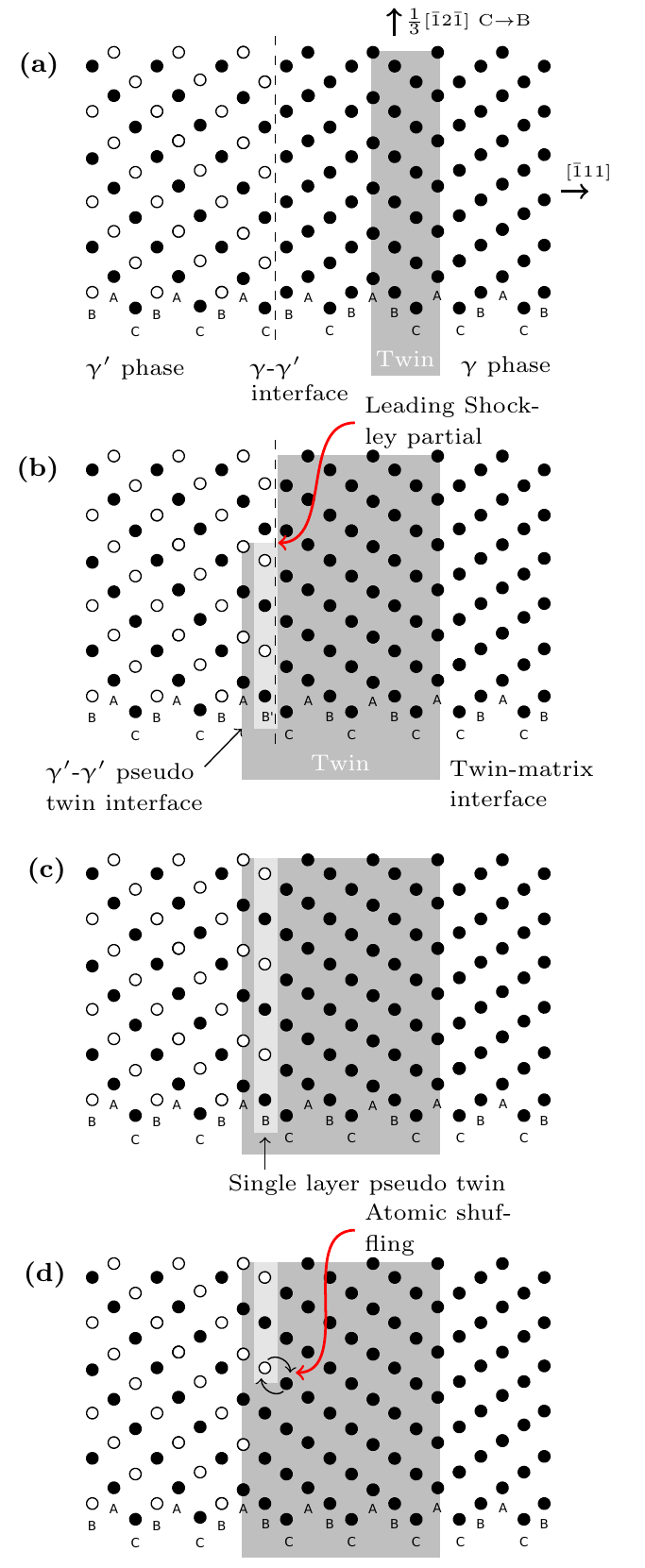} 
	\caption{Schematic showing formation of a single layer pseudo twin at a $\upgamma-\upgamma'$ interface and consequent precipitate dissolution. (a) Twin within the $\upgamma$ matrix, dark grey region. (b) Twin growth into the first $\upgamma'$ atomic layer showing creation of a single layer pseudo twin in the wake of the leading Schockley partial, light grey region. (c) The partial has made its way across a narrow stretch of the $\upgamma-\upgamma'$ interface leaving behind it a single layer pseudo twin. A parent-twin $\upgamma'$-psuedo twin interface is shown on the left and a pseudo twin matrix on the right. (d) Atomic shuffling in the wake of the leading Shockley partial disorders the single layer pseudo twin slowly dissolving the $\upgamma'$ precipitate during the process of twin growth.}
	\label{fig:schem}
\end{figure}

As the twin boundary moves by the motion of $\frac{1}{3}$$<$$112$$>$ partials gliding on adjacent $\{$$111$$\}$ planes, the twin thickens one atomic plane at a time, Figure~\ref{fig:schem}(a). If the twin is locally contained within a region of disordered $\upgamma$ phase there will come a time during the thickening process were the twin boundary will interact with a $\upgamma / \upgamma'$ interface, Figure~\ref{fig:schem}(b). The $\frac{1}{3}$$<$$112$$>$ partial is a $\frac{1}{6}$$<$$112$$>$ superlattice partial in $\upgamma'$. Shearing of the first $\{111\}$ $\upgamma'$ atomic layer by the superlattice partial creates a single layer pseudo-twin within the $\upgamma'$ precipitate. If this pseudo-twin was in the bulk of the precipitate atomic reordering with atoms from an adjacent $\{111\}$ plane would convert it to a true twin structure. However, in the current case we are presented with a single layer pseudo-twin boundary on one side of the planar fault and a $\upgamma / \upgamma'$ interface at the other, Figure~\ref{fig:schem}(c). We hypothesise that such a configuration is thermodynamically unstable at the testing temperature of \si{800\celsius} and so the single layer pseudo-twin would disorder, via atomic reshuffling with the $\upgamma$ phase, dissolving into the $\upgamma$ matrix, Figure~\ref{fig:schem}(d). This results in a $\upgamma / \upgamma'$ twin boundary as observed in the present work.

Such a mechanism is important for two reasons. (i) The layer-by-layer dissolution process of the ordered precipitate phase at the $\upgamma / \upgamma'$ twin interface may be a rate limiting step for twin thickening. (ii) The formation of precipitate free bands which traverse entire $\upgamma$ grains can provide resistance-free dislocation paths, which, in turn increase creep rates and result in large strains at grain boundaries. This may be more prevalent during a transition to a high stress low temperature creep regime where twin motion is not favoured and dislocations glide free from the resistance of ordered precipitates along previously formed twins.

\section{Summary and conclusions}
The tensile creep performance of a polycrystalline Co/Ni-base superalloy processed so as to produce a multimodal $\upgamma'$ distribution has been examined at \si{800\celsius} and \si{300\,\mega\pascal}. The main findings are as follows:
\begin{itemize}
	\item The rupture life of the alloy is similar to that of commercial alloy RR1000 tested under the same conditions.
	\item \ce{M23C6} carbides nucleate at $\upgamma$ grain boundaries during creep. These carbides observe a cube-cube orientation relationship with the grain which they share a semi-coherent interface with, while acting as defect nucleation sites in the adjacent grain with which they share an incoherent interface.
	\item $\upgamma'$ precipitate-free twins are observed to cross large regions of $\upgamma$ grains. Solute segregation at the $\upgamma$/$\upgamma'$ phase boundary is suggested to result in the dissolution of the precipitate as the twin grows from the dissordered matrix phase into the ordered precipitate, slowly consuming the precipitate and creating a twin predominantly free of the ordered $\upgamma'$ phase. This mechanisms is similar to the Kolbe re-ordering with the exception that in the classical mechanism the pseudo twin is contained within the ordered precipitate whereas in the present case it is sandwiched between the ordered precipitate and the disordered matrix phase.
\end{itemize}

These findings suggest that polycrystalline Co/Ni-base superalloys have the potential to compete with commercial Ni-base superalloys in terms of creep performance. Precipitation of \ce{M23C6} carbides act as defect nucleation sites at the incoherent carbide-matrix interface, and that the composition of the V208C Co/Ni-base superalloy is prone to precipitate dissolution during twin growth when creep tested in tension at \si{800\celsius} and \si{300\,\mega\pascal}.



\section*{Acknowledgement}
This work was funded under the Rolls-Royce-EPSRC strategic partnership in structural metallic systems for gas turbines (EP/M005607/1).
The authors would like to thank Dr Catrin Davies and Alex Toth from the Mechanical Engineering Department at Imperial College London for help with the use of the creep rigs within the High Temperature Testing laboratory. Our thanks is also extended to Mike Lennon and Ben Wood for their assistance in creep specimen preparation. 
VAV would like to acknowledge support from Rolls-Royce plc and Imperial College London under the Imperial College Research Fellowship scheme.

\section*{References}
\bibliography{./paper1_creep_v09.bbl}
\bibliographystyle{elsarticle-num}



\end{document}